\documentclass[10pt,twocolumn,english,prd,superscriptaddress,nofootinbib,preprintnumbers,showpacs,floatfix]{revtex4-2}
\usepackage{amsmath,amssymb,amsthm,hyperref}
\usepackage[utf8]{inputenc}
\usepackage{graphicx}
\usepackage{xcolor}

\newcommand{\bdm}{\begin{displaymath}}
\newcommand{\edm}{\end{displaymath}}

\usepackage{orcidlink}

\begin{document}

\title{ A regularized photon-pair geometry motivated by the ER=EPR conjecture}

\author{Kimet Jusufi\,\orcidlink{0000-0003-0527-4177}}
\email{kimet.jusufi@unite.edu.mk}
\affiliation{Physics Department, State University of Tetovo, Ilinden Street nn, 1200, Tetovo, North Macedonia}

\author{Francisco S.N. Lobo\,\orcidlink{0000-0002-9388-8373}}%
\email{fslobo@fc.ul.pt}
\affiliation{Departamento de F\'{i}sica, Faculdade de Ci\^{e}ncias da Universidade de Lisboa, Campo Grande, Edif\'{\i}cio C8, P-1749-016 Lisbon, Portugal}
\affiliation{Instituto de Astrof\'{\i}sica e Ci\^{e}ncias do Espaco, Faculdade de Ci\^encias da Universidade de Lisboa, Campo Grande, Edif\'{\i}cio C8, P-1749-016 Lisbon, Portugal}%

\author{Douglas Singleton\,\orcidlink{0000-0001-9155-7282}}
\email{dougs@mail.fresnostate.edu}
\affiliation{Physics Department, California State University, Fresno, CA 93740}

\date{\today}

\begin{abstract}
We regularize the Aichelburg--Sexl shock-wave geometry for massless
particles by smearing the point-like source over a string-inspired
length scale $l_0$.  A radial extension of the transverse geometry
admits a zero-throat Einstein--Rosen interpretation.  For a photon
wave packet of longitudinal extent $L$, the regularized gravitational
self-energy of a transversely separated pair is
\(
E_{\rm GSE}\sim
\frac{4G(\hbar\omega)^2}{c^4L}
\ln\!\left(\frac{d^2}{l_0^2}\right),
\)
where $d$ is the transverse separation.  The factor $1/L$ strongly
suppresses the interaction, giving gravitational stability times
greater than $10^{30}$ years for optical photons.
We also examine an effective two-dimensional entanglement-entropy
description of the shock-wave geometry.  The entropy construction
reproduces the same dependence on the scales $d$, $L$, and $l_0$, and
can be calibrated to reproduce the normalization of the direct
self-energy calculation.  However, we find that the gravitationally
induced excess entropy, and hence the self-energy inferred from this
prescription, is the same for Bell-entangled and unentangled photon
pairs having identical energy-momentum distributions.  This agrees
with the direct geometrical calculation, since the classical
two-shock metric depends on the stress-energy tensor but not on the
polarization entanglement.
The present construction should therefore be understood as a
semiclassical ER-like geometry motivated by ER=EPR, rather than as a
complete realization of the conjecture.  A geometry whose
connectivity tracks the amount of quantum entanglement would require
a quantum-gravitational description sensitive to state-dependent
correlations beyond the one-point stress-energy tensor.
	
\end{abstract}

\maketitle

\section{Introduction}

Quantum mechanics provides an extraordinarily successful description of microscopic physics, accurately accounting for phenomena such as superposition and quantum entanglement. At the foundation of the theory lies the Schr\"odinger equation \cite{Schrodinger:1926gei}, which governs the unitary evolution of quantum states. Despite its empirical success, however, the interpretation of quantum measurement and the physical origin of wave-function collapse remain unresolved. An interesting approach to addressing this problem is provided by gravity-induced collapse models, originally proposed by Di\'osi and Penrose \cite{Diosi:1986nu,Diosi:1984wuz,Penrose:1996cv,Penrose:1998dg} and further developed in a variety of contexts \cite{Gasbarri:2017alx,Tomaz:2024rzu,Bose:2024nhv,Trillo:2025kio,Gundhi:2025aaa,Diosi:2022kay,Donadi:2020kzc,Hossenfelder:2025uqw}. In the Di\'osi--Penrose (DP) framework, the superposition of different mass configurations corresponds to a superposition of distinct spacetime geometries. Since general relativity associates gravity directly with spacetime curvature, such geometries cannot be simultaneously well defined, leading to an intrinsic instability of the quantum state. Penrose argued that this tension between the equivalence principle and quantum linearity provides a natural mechanism for spontaneous wave-function reduction \cite{Penrose:2014nha,Howl:2018qdl}. The collapse timescale is then determined by the gravitational self-energy associated with the difference between the competing geometries.

Most investigations of gravitationally induced collapse have focused on massive systems, where the gravitational self-energy can become appreciable. By contrast, the role of massless particles, particularly photons, has received considerably less attention. Although photons possess no rest mass, they carry energy and momentum and therefore gravitate through the stress-energy tensor. Understanding whether quantum superpositions of photons are also subject to gravity-induced instability is therefore of both conceptual and phenomenological importance, especially in view of recent developments connecting quantum information, spacetime geometry, and entanglement.

In this work we investigate gravity-induced collapse for photons by employing a regularized version of the Aichelburg--Sexl (AS) shock-wave geometry. The regularization is motivated by nonlocal modifications of gravity and string-inspired T-duality arguments \cite{Sathiapalan:1986zb,Padmanabhan:1996ap,Smailagic:2003hm,Fontanini:2005ik}, which imply the existence of a fundamental zero-point length $l_0 \sim \sqrt{\alpha'}$.
Such a minimal length effectively smears point-like sources and removes curvature singularities, leading to finite gravitational potentials at short distances. Regularized geometries of this type have been extensively explored in the context of black holes, cosmology, and nonlocal gravity models \cite{Nicolini:2022rlz,Nicolini:2019irw,Jusufi:2024dtr,Jusufi:2025qgd}. More recently, nonlocal gravitational effects associated with zero-point length deformations have been incorporated directly into spacetime geometry \cite{Jusufi:2025qgd}. These ideas have been applied to the construction of Einstein--Rosen bridge configurations with implications for the ER=EPR conjecture \cite{Jusufi:2025rlr}, as well as to modified Schr\"odinger--Newton equations and gravity-induced collapse scenarios for massive particles \cite{Jusufi:2025lrq}. In parallel, recent work has suggested that ER=EPR-type effects could in principle be constrained through atomic systems such as hydrogen \cite{Javed:2025hvr}. Interestingly, the zero-throat wormhole configurations emerging in regularized nonlocal geometries \cite{Jusufi:2025lrq} appear to evade such constraints. It is useful to clarify why the constraints of
Ref.~\cite{Javed:2025hvr} do not apply to the present
construction. Their analysis assumes that the ER=EPR
wormhole possesses a finite effective cross-sectional area,
allowing a fraction of the electric flux surrounding an
entangled charged particle to leak into the wormhole. The
resulting modification of the electromagnetic field leads to
observable corrections to the hydrogen spectrum. By
contrast, the regularized Aichelburg--Sexl geometry studied
here possesses a zero-throat Einstein--Rosen bridge. Indeed,
Ref.~\cite{Javed:2025hvr} explicitly notes that for the
zero-throat proposal of Ref.~\cite{Jusufi:2025rlr}
there would be no electric-field leakage and therefore no
corresponding hydrogen constraint. Consequently, the
bounds derived in Ref.~\cite{Javed:2025hvr} are not
expected to apply to the present model.

Motivated by these developments, we extend the Di\'osi--Penrose framework to photon states propagating in regularized AS geometries. Our analysis explores how the introduction of a zero-point length modifies the gravitational self-energy of photon superpositions and alters the associated collapse timescale. In particular, we investigate whether nonlocal geometric effects can provide a natural ultraviolet regularization of gravitational self-interaction while preserving the essential features of Penrose-type collapse mechanisms. Beyond its implications for foundational aspects of quantum mechanics, this approach may also shed light on possible connections between nonlocal gravity, emergent spacetime geometry, and quantum entanglement.

We construct a regularized AS metric whose extended transverse geometry admits a zero-throat Einstein--Rosen interpretation.  This provides an ER-like geometric setting in which to explore photon pairs motivated by the ER=EPR conjecture.  A key ingredient of our analysis is the explicit identification of the photon longitudinal extent $L$, which must appear in the self-energy functional for dimensional consistency.  The resulting interaction energy is suppressed by a factor $1/L$ and is therefore far weaker than a point-particle estimate would suggest.

It is important, however, to distinguish the result obtained here from a complete realization of ER=EPR.  The regularized geometry is determined by the classical stress-energy tensor of the photons. Consequently, two photon pairs with the same energies, momenta, and wave-packet profiles generate the same two-shock geometry whether their polarization state is Bell entangled or separable.  The present semiclassical metric therefore captures an ER-like geometry but does not by itself encode the EPR entanglement or its magnitude.

For an isolated optical photon the corresponding gravitational stability time surpasses $10^{30}$ years.  For a back-to-back pair created in $e^+e^-\rightarrow 2\gamma$, the regularized interaction energy emerges as the two wave packets separate transversely.  The same interaction energy is obtained for an unentangled pair with the same stress-energy distribution.  Thus the self-energy is a property of the null energy distributions in the present semiclassical model, not a measure of polarization entanglement.

This limitation is especially relevant to the ER=EPR interpretation. An ideal quantum-nondemolition measurement may destroy the polarization entanglement while leaving the photon energies and momenta, and hence the classical stress-energy tensor, essentially unchanged.  The classical two-shock geometry would then remain unchanged.  Establishing a bridge whose existence or connectivity changes with the quantum state would require a description sensitive to higher-order stress-tensor correlations, the density matrix, or other genuinely quantum-gravitational data.

We further examine an effective entanglement-entropy interpretation of the gravitational self-energy.  The entropy of null intervals straddling the shock geometry contains the same scales $d$, $L$, and $l_0$ that appear in the direct gravitational calculation.  With an effective entanglement temperature, the prescription can be calibrated to reproduce the direct self-energy.  Carrying the same calculation through for a factorized photon state shows, however, that the gravitationally induced excess entropy is identical to that of the Bell-entangled state when the classical stress-energy distributions coincide.  This agreement clarifies both the usefulness and the limitations of the entropy construction: it supplies an information-theoretic description of the shock geometry, but it does not make the geometry depend on the polarization entanglement.

This work thus establishes a theoretical laboratory for exploring quantum gravity effects in the massless sector, revealing a deep suppression of gravitational phenomena for ultra-relativistic/massless wave packets.

The paper is organized as follows. In Section~\ref{SecII} we introduce the regularized Aichelburg--Sexl metric, showing how the string-inspired scale $l_0$ removes the curvature singularity and how a radial extension admits a zero-throat Einstein--Rosen interpretation. In Section~\ref{SecIII} we derive the gravitational self-energy functional for a photon wave packet of finite longitudinal extent. The corresponding gravitational stability time is estimated in Section~\ref{SecIV}. Section~\ref{SecV} applies the regularized two-shock geometry to photon pairs created in electron--positron annihilation and clarifies the extent to which the construction can be interpreted within ER=EPR. In Section~\ref{SecVI} we compare the entanglement-entropy construction for Bell-entangled and unentangled photon pairs and relate both results to the direct geometrical self-energy. Finally, we conclude in Section~\ref{Sec-VII}.

\section{Regularized Aichelburg--Sexl Metric}\label{SecII}

The standard Aichelburg--Sexl metric describes the gravitational field of an ultrarelativistic massless particle \cite{Aichelburg:1970dh}:
\begin{equation}
	ds^2 = -du\,dv + dx^2 + dy^2 + \Phi(x,y)\,\delta(u)\,du^2,
	\label{ASmetric}
\end{equation}
in the unregularized case $ \Phi(\rho) = -\,(8\,G E/c^4)\,\ln(\rho/\rho_0)$
(with $\rho = \sqrt{x^2 + y^2}$), $E$ is the energy, and $u=ct-z$ and $v=ct+z$ are light cone coordinates. The logarithmic potential produces a curvature singularity at $\rho=0$.

To regularize this, we introduce a smeared transverse energy density
\begin{equation}
	\rho_E(\rho) = \frac{E\,l_0^2}{\pi\,(\rho^2 + l_0^2)^2}.
	\label{rhoE}
\end{equation}
For a $pp$-wave, the Einstein equations reduce to a two-dimensional Poisson equation in the transverse plane:
\begin{equation}
	\Delta_\perp \Phi(\rho)
	=\frac{1}{\rho}\,\frac{d}{d\rho}\!\left(\rho\,\frac{d\Phi(\rho)}{d\rho}\right)
	= -\,\frac{16\pi G}{c^4}\,\rho_E(\rho).
	\label{Poisson}
\end{equation}

We adopt the regularized potential
\begin{equation}
	\Phi(\rho) = -\,\frac{4\,G E}{c^4}\,
	\ln\!\left(1 + \frac{\rho^2}{l_0^2}\right).
	\label{Phi}
\end{equation}
This is finite at $\rho=0$ ($\Phi(0)=0$) and asymptotically behaves as
$\Phi(\rho) \approx -\,\dfrac{8\,G E}{c^4}\,\ln(\rho/l_0)$ for $\rho \gg l_0$
(up to an additive constant). Its Laplacian is
\begin{equation}
	\Delta_\perp \Phi = -\,\frac{16\,G E\,l_0^2}{c^4\,
			(\rho^2 + l_0^2)^2},
	\label{laplacian}
\end{equation}
which exactly reproduces the source \eqref{rhoE} when inserted into \eqref{Poisson}.
The density \eqref{rhoE} is properly normalized:
\begin{equation}
	\int_0^\infty \rho_E(\rho)\,2\pi\rho\,d\rho = E,
\end{equation}
and reduces to a delta function as $l_0 \to 0$.
Finally, the regularized Aichelburg--Sexl metric in polar coordinates reads
\begin{equation}
	ds^2 = -du\,dv + d\rho^2 + \rho^2 d\theta^2
	+ \Phi(\rho)\,\delta(u)\,du^2.
	\label{ASmetric1}
\end{equation}
The metric in \eqref{ASmetric1} describes a massless particle whose shock is localized on the null surface $u=0$ and which propagates in the $+z$ direction. In a similar way, the AS metric for a photon moving in the $-z$ direction is obtained by exchanging $u$ and $v$, leading to
\begin{equation}
	ds^2 = -du\,dv + d\rho^2 + \rho^2 d\theta^2
	+ \Phi(\rho)\,\delta(v)\,dv^2.
	\label{ASmetric2}
\end{equation}

To model a back‑to‑back pair -- one photon travels in the $+z$ direction and the other in the $-z$ direction -- we must superpose the shocks on the two distinct null surfaces $u=0$ and $v=0$. In the weak‑field approximation, the metric perturbation is linear, so the combined geometry is given by summing Eqs. \eqref{ASmetric1} and \eqref{ASmetric2} giving
\begin{equation}
	ds^2
	=
	-du\,dv
	+d\rho^2+\rho^2 d\theta^2
	+\Phi(\rho) \left( \delta(u)\,du^2
	+\delta(v)\,dv^2 \right),
	\label{superposed}
\end{equation}
with the regularized potential, $\Phi (\rho)$, given by \eqref{Phi}.

To make the ER bridge nature of the metric in \eqref{superposed} more apparent we introduce the radial coordinate transformation in the metric \eqref{ASmetric1} as follows $r^2 = \rho^2 + l_0^2$, which gives
\begin{equation}
	\rho = \sqrt{r^2 - l_0^2}, \quad r \in (-\infty, -l_0] \cup [l_0, +\infty).
	\label{rcoord}
\end{equation}
The transverse metric \eqref{ASmetric1} then becomes that of a two-dimensional wormhole:
\begin{equation}
	d\Sigma^2 = \frac{dr^2}{1 - l_0^2/r^2} + (r^2 - l_0^2) d\theta^2.
	\label{transverse}
\end{equation}
As $|r| \to \infty$, this approaches flat Euclidean space $dr^2 + r^2 d\theta^2$. The throat of the wormhole lies at $|r| = l_0$, where the geometry remains completely regular when expressed in the original $\rho$ coordinates. Note, this is a zero-throat wormhole, since the throat radius is zero at $|r| = l_0$. The full regularized metric for the back-to-back photon pair therefore reads
\begin{equation}
	\begin{aligned}
		ds^2 &= -du\,dv
		+ \frac{dr^2}{1 - l_0^2/r^2}
		+ (r^2 - l_0^2)\,d\theta^2\\
		&\quad - \frac{8G E_\gamma}{c^4}\,
		\ln\!\Bigl|\frac{r}{l_0}\Bigr|\,
		\left(\,\delta(u)\,du^2 + \delta(v)\,dv^2\,\right).
	\end{aligned}
	\label{wormholeBB}
\end{equation}
The shock on $u=0$ is the $+z$-traveling photon, and the shock on $v=0$ the $-z$-traveling photon. The two asymptotic branches $r\to+\infty$ and $r\to-\infty$ admit an Einstein--Rosen interpretation for the back-to-back photon pair. At this classical level, however, the geometry is determined only by the null stress-energy distributions and is therefore the same for entangled and unentangled polarization states having identical energies and momenta.

We note that the back‑to‑back metric~\eqref{wormholeBB} with two shock waves moving in opposite directions is {\it not} an exact solution of the full Einstein equations; the superposition is justified only in the linearized approximation, where the gravitational fields of the two photons do not interact. For the purpose of computing the leading-order gravitational self-energy, this approximation is sufficient.  Its use in connection with ER=EPR is necessarily interpretive, since the classical metric does not contain a parameter measuring the photon entanglement.

\section{Photon Wave Packet with Gravitational Self-Energy}\label{SecIII}

For a photon wave packet of finite longitudinal extent $L$, we make the crude, but allowable approximation that the energy, $E=\hbar \omega$, is distributed over the null direction $u$ with constant linear density $\mu \equiv E/L = \hbar\omega/L$. 
The stress‑energy tensor takes the form $T_{uu}(\mathbf{x},u) = \,
\frac{\rho_E(\mathbf{x}_\perp)}{ L}$ over $0<u<L$, where the energy density is normalized as $\int \rho_E\,d^2x_\perp = \hbar \omega$. Consequently, the two‑dimensional Poisson equation that determines the
gravitational potential at a given $u$ is
	\[
	\Delta_\perp \Phi(\mathbf{x}_\perp)
	= -\,\frac{16\pi G}{c^4}\,\frac{\rho_E(\mathbf{x}_\perp)}{L} .
	\]

For a photon wave packet of finite longitudinal extent $L$, the strict delta-function AS shock profile is replaced by an effective longitudinally averaged $pp$-wave profile. For simplicity, we continue to denote the corresponding transverse gravitational potential by $\Phi ({\bf x}_\perp )$.
In the regularized AS metric, the gravitational potential generated by this linear energy density is the 2D convolution
\begin{equation}
	\Phi(\mathbf{x}_\perp)
	= -\,\frac{4G}{c^4 L}
	\int \rho_E(\mathbf{y}_\perp)\,
	\ln\!\left(1 + \frac{|\mathbf{x}_\perp - \mathbf{y}_\perp|^2}{l_0^2}\right)
	d^2 y_\perp ,
	\label{potential2D}
\end{equation}
where the prefactor $-4G/(c^4 L)$ follows from the regularized
potential $\Phi(\rho) = -\,\frac{4G\hbar \omega }{Lc^4}\,\ln(1+\rho^2/l_0^2)$.  The potential is now explicitly of order $1/L$, reflecting the fact that only a fraction of the total energy
contributes to any single $u$‑slice.
The logarithm is regularized at short distances by the T-duality-inspired
zero-point length $l_0$, so $\Phi(0)=0$ and the potential remains free of
singularities. For large separations ($|\mathbf{x}_\perp - \mathbf{y}_\perp|
\gg l_0$) it recovers the familiar logarithmic behavior up to an additive constant.

The gravitational self-energy per unit longitudinal length is
\begin{equation}
	\mathcal{E}^{\rm GSE}(u)
	= -\frac{1}{2} \int \left( \frac{\rho_E(\mathbf{x}_\perp)}{L}\right)\,
	\Phi(\mathbf{x}_\perp)\,d^2x_\perp .
	\label{selfenergyperlength}
\end{equation}
Substituting the expression for $\Phi$ and using the regularized kernel gives
\begin{eqnarray}
	\mathcal{E}^{\rm GSE}
	= &&\frac{2G}{c^4 L^2}
	\iint \rho_E(\mathbf{x}_\perp)\,\rho_E(\mathbf{y}_\perp) \nonumber \\
    &&\ln \left( 1 + \frac{|\mathbf{x}_\perp - \mathbf{y}_\perp|^2}{l_0^2} \right)
	d^2x_\perp d^2y_\perp,
	\label{GSEdouble}
\end{eqnarray}
where $d^2a_\perp = d^2x_\perp\,d^2y_\perp$.
The total gravitational self‑energy $E^{\rm GSE}$ of the photon
wave packet is obtained by integrating $\mathcal{E}^{\rm GSE}$ over its longitudinal extent. Since $\mathcal{E}^{\rm GSE}$ is independent of $u$, for a quasi‑monochromatic packet of length $L$ we have
\begin{align}
	E^{\rm GSE}
		=& \int_0^L \mathcal{E}^{\rm GSE}(u)\,du
		= L \mathcal{E}^{\rm GSE},
\end{align}
yielding 
\begin{eqnarray}
	E^{\rm GSE}
		= && \frac{2G}{c^4 L}
		\iint \rho_E(\mathbf{x}_\perp)\rho_E(\mathbf{y}_\perp)\nonumber \\
        &&\ln \left( 1 + \frac{|\mathbf{x}_\perp - \mathbf{y}_\perp|^2}{l_0^2} \right)
		d^2x_\perp d^2y_\perp.
		\label{Etotal}
\end{eqnarray}

\section{Collapse Time for a Photon Superposition}\label{SecIV}

Inspired by the Di\'osi--Penrose framework \cite{Diosi:1984wuz,Diosi:1986nu,Penrose:1996cv,Penrose:1998dg}, we estimate the characteristic gravitational stability timescale using the regularized
gravitational self-energy derived in the previous section. In this sense, standard quantum mechanics without gravitational effects should be viewed only as an approximation theory. In the case of photons, as we shall argue, such effects are small, and we can assume the validity of the superposition principle. In particular, we aim to demonstrate this by using the collapse time of the photon's wave function due to gravitational self-interaction. 

Consider a photon in a transverse spatial superposition
\begin{equation}
	| \psi_\gamma \rangle   \approx \frac{1}{\sqrt{2}} \Big( |\text{packet at }\mathbf{r}_1\rangle + |\text{packet at }\mathbf{r}_2\rangle \Big),
\end{equation}
with separation $d = |\mathbf{r}_1 - \mathbf{r}_2| \gg l_0$. In the point-particle approximation for the transverse energy distributions ($l_0 \ll d$), we can write 
\begin{equation}
	\rho_E(\rho) = \lim_{l_0 \to 0} E \left(\frac{ l_0^2}{\pi (\rho^2 + l_0^2)^2}\right) = E \delta^{(2)}(\rho).
	\label{rhoEapprox}
\end{equation}

With this we can define the transverse energy density as the sum of the two localized packets, each carrying energy $\hbar \omega$, which leads to
\begin{equation}
\rho_E^{\rm tot}(\mathbf{x}_\perp) = \hbar\omega \left[ \delta^{(2)}(\mathbf{x}_\perp - \mathbf{r}_1) + \delta^{(2)}(\mathbf{x}_\perp - \mathbf{r}_2) \right],
\label{perp-x}
\end{equation}
where $d = |\mathbf{r}_1 - \mathbf{r}_2| \gg l_0$. We can obtain $\rho_E^{\rm tot}(\mathbf{y}_\perp)$ from \eqref{perp-x} by letting $\mathbf{x}_\perp \to \mathbf{y}_\perp$.

We now use Eq.~\eqref{Etotal} to calculate the total gravitational self-energy for a wave packet of longitudinal extent $L$ by evaluating the double integral with the delta‑function densities $\rho_E^{\rm tot}(\mathbf{x}_\perp) \rho_E^{\rm tot}(\mathbf{y}_\perp)$. This gives
\begin{equation}
	E^{\rm GSE}
	= \frac{4G (\hbar\omega)^2}{c^4\,L}\,
	\ln\!\left(1+\frac{d^2}{l_0^2}\right)
	\approx \frac{4G (\hbar\omega)^2}{c^4\,L}\,
	\ln\!\left(\frac{d^2}{l_0^2}\right).
	\label{deltaE}
\end{equation}
The factor $1/L$ is essential for dimensional consistency; it converts the energy‑per‑unit‑length of the shock‑wave geometry into a physical energy.
	For a quasi‑monochromatic optical photon we may take $L$ to be the wavelength
	$L \simeq \lambda = 2\pi c/\omega$, leading to
\begin{equation}
	E^{\rm GSE} \simeq
		\frac{2G\,\hbar^2\omega^3}{\pi c^5}\,
		\ln\!\left(\frac{d^2}{l_0^2}\right).
\end{equation}
The quantity $E_{\rm GSE}$ is the interaction energy associated with the two separated null energy distributions, rather than the total energy $2\hbar\omega$ carried by the photons.  Within an ER-like interpretation, it may be associated with the gravitational binding of the extended geometry.  The calculation itself, however, depends only on the energy distributions and gives the same result for entangled and unentangled photon pairs with identical stress-energy tensors.

Following Penrose, the characteristic collapse time is
\begin{equation}
	\tau_{\rm collapse} \sim \frac{\hbar}{E^{\rm GSE}}
	\approx \frac{\hbar\,c^4\,L}{4G (\hbar\omega)^2\ln(d^2/l_0^2)}.
	\label{tauexact}
\end{equation}
Inserting $L = 2\pi c/\omega$ gives the explicit frequency scaling
\begin{equation}
	\tau_{\rm collapse} \sim
		\frac{\pi\,c^5}{2G\,\hbar\,\omega^3\,\ln(d^2/l_0^2)}.
\end{equation}

For an optical photon ($\hbar \omega \approx 2.5\,\text{eV} \approx 4 \times 10^{-19}\,\text{J}$), laboratory separation $d \sim 1\,\text{m}$, and $l_0$ near the Planck scale ($\sim 10^{-35}\,\text{m}$), the large logarithm is overpowered by the tiny $(\hbar \omega)^3$ and the enormous $c^5/G$, yielding $\tau_{\rm collapse}$ on the order of $10^{30}$--$10^{40}$ years or longer---vastly exceeding the age of the universe ($\sim 1.4 \times 10^{10}$ yr). Thus, isolated photons are effectively immune to gravitational self-collapse. 

For optical photons ($\hbar \omega \sim$ eV), the collapse time exceeds $10^{30}$ years. Even for gamma rays near the electron–positron annihilation scale ($\sim 511$ keV), the collapse time remains astronomically long ($\sim 10^{14}$ years). Only at extremely high energies approaching the TeV–PeV regime would the collapse time become phenomenologically short, though in that regime the approximations underlying the present treatment may become unreliable.
We emphasize that the use of the characteristic timescale
$\tau_{\rm collapse}\sim\hbar/E_{\rm GSE}$ in the present work should not
be identified with the original stochastic Di\'osi--Penrose (DP)
collapse model. The natural parameter-free version of the DP model has
been experimentally excluded by the underground search for
collapse-induced radiation reported in Ref.~\cite{Donadi:2020kzc}. In the
present work, however, we use the Penrose gravitational self-energy only
as a motivation for estimating a characteristic gravitational stability
timescale based on the regularized Aichelburg--Sexl self-energy of a
photon wave packet. We do not introduce the universal stochastic noise
field responsible for dynamical collapse in the original DP model, nor do
we calculate collapse-induced diffusion or spontaneous radiation.
Consequently, the present calculation should be regarded as a
deterministic gravitational stability estimate rather than an
implementation of the original stochastic Di\'osi--Penrose collapse
mechanism. Therefore, the experimental constraints on the original
parameter-free DP model do not directly apply to the framework developed
here.

\section{Photon-Pair Geometry and Its Relation to ER=EPR}
\label{SecV}

The process $e^+e^-\rightarrow 2\gamma$ provides a clean source of back-to-back photon pairs.  In the low-energy center-of-momentum frame, the photons are emitted with equal energies $E_\gamma=m_ec^2\simeq511~{\rm keV}$.  When the initial electron--positron state has total angular momentum zero, the photon polarizations may be described by the Bell state
\begin{equation}
\label{bell-state}
 |\psi\rangle =
 \frac{1}{\sqrt{2}}
 \left(
 |H\rangle_1|V\rangle_2
 -
 |V\rangle_1|H\rangle_2
 \right),
\end{equation}
where $|H\rangle$ and $|V\rangle$ denote orthogonal linear
polarizations.

The ER=EPR conjecture motivates interpreting quantum entanglement in terms of a nontraversable Einstein--Rosen bridge.  The regularized two-shock metric constructed in Sec.~\ref{SecII} provides an ER-like geometry for the two oppositely propagating null energy distributions.  We emphasize, however, that this classical geometry is not selected by the Bell state.  A separable polarization state with the same photon energies, momenta, and wave-packet profiles produces the same stress-energy tensor and therefore the same metric.  The construction should consequently be viewed as a semiclassical geometry compatible with an ER=EPR interpretation, rather than as a derivation of the state-dependent equality between ER and EPR.

As the two photon wave packets propagate away from their creation region, their transverse separation may be represented by $d=|\mathbf r_1-\mathbf r_2|$.  Repeating the calculation of Sec.~\ref{SecIV} for two separated null energy densities gives the logarithmic interaction energy in Eq.~\eqref{deltaE}.  This result applies to any two photon wave packets with the same energy distributions, independently of whether their polarization state is entangled.  Thus the separation dependence characterizes the regularized gravitational interaction of the two shocks; calling it the stretching energy of an ER bridge is an additional interpretation motivated by ER=EPR.

The vast majority of the total energy, $2\hbar\omega$, remains in the photon wave packets.  The interaction energy is extremely small in comparison.  For a Bell-entangled pair, the extended geometry may be interpreted as an ER-like counterpart of the EPR correlations. Nevertheless, the present semiclassical calculation does not show that the bridge disappears when those correlations are removed.

We can use Eq.~\eqref{deltaE} to perform a consistency check. When the separation of the two photon packets approaches the scale, $d \to l_0$, the gravitational self‑energy $E^{\rm GSE}$ from Eq. \eqref{deltaE} reads 
\begin{equation}
	E^{\rm GSE}
	= \frac{4G (\hbar\omega)^2}{c^4\,L} \ln(2)~.
	\label{deltaE2}
\end{equation}
If, in addition, $l_0$ is identified with the Planck length, $L\sim l_0$, and the photon energy is taken to be of Planck order, $\hbar\omega\sim\hbar c/l_0$, then Eq.~\eqref{deltaE2} gives, up to a numerical factor of order unity,
\begin{equation}
	t_{\rm collapse} \sim \frac{\hbar}{E^{\rm GSE}}
	\sim \frac{l_0}{c}  \sim t_{Pl},
	\label{tcollapse}
\end{equation}
Geometrically, this corresponds to a ``zero‑throat'' Einstein--Rosen bridge. Indeed, this behavior is consistent with the independent analysis of regularized wormholes carried out in~\cite{Jusufi:2025rlr}. In that work, it was demonstrated that the exotic matter required to sustain a traversable wormhole also vanishes when the throat size reduces to the same fundamental length scale $l_{0}$, thereby satisfying the null energy condition. 
The appearance of the same fundamental scale $l_0$ in the self-energy estimate and in the regularized wormhole construction supports the internal consistency of the framework.  The two results should not, however, be identified directly: Eq.~\eqref{deltaE2} remains finite and nonzero at $d=l_0$, whereas Ref.~\cite{Jusufi:2025rlr} concerns the behavior of the exotic matter supporting a distinct regularized wormhole geometry.

The zero-throat nature of the present construction also clarifies its
relation to the hydrogen constraints derived in
\cite{Javed:2025hvr}. Those authors assumed that an ER=EPR
wormhole possesses a finite effective cross-sectional area, allowing part
of the electric flux surrounding an entangled charged particle to leak
into the wormhole, thereby modifying the hydrogen spectrum. As explicitly
noted in \cite{Javed:2025hvr}, this mechanism is absent for the
zero-throat proposal of Ref.~\cite{Jusufi:2025rlr}, since a
vanishing throat area implies that no electric flux enters the wormhole.
Consequently, the hydrogen constraints of Ref.~\cite{Javed:2025hvr}
are not expected to apply to the regularized Aichelburg--Sexl wormhole
considered here.

The present construction also differs from the classical finite-throat wormhole scenarios discussed in \cite{Dai:2020ffw}. In that work, observable consequences arise from the finite spatial extent and dynamical properties of the wormhole geometry. By contrast, the regularized Aichelburg--Sexl wormhole constructed here possesses a zero throat and its observable consequences arise through the regularized gravitational self-energy and the associated entanglement entropy rather than through propagation effects associated with a finite wormhole throat. Thus the observational constraints discussed in Ref.~\cite{Dai:2020ffw} do not directly apply to the present construction.

Finally, our model should also be compared with the recent
Einstein--Dirac--Maxwell construction of \cite{Kain:2023ore}, which
provides a concrete classical spacetime realization of ER=EPR for two
entangled spin-$1/2$ particles. In that work, the matter sector is
described by Dirac wave functions coupled to classical gravity and a
classical gauge field, and numerical evolution shows that the wormhole
becomes nontraversable through black-hole formation. A particularly
interesting feature of Ref.~\cite{Kain:2023ore} is that part of the
wormhole throat shrinks dynamically, bringing the two particle
distributions into close proximity inside the bridge. The present
construction is complementary: instead of a finite-throat
Einstein--Dirac--Maxwell wormhole for massive fermions, we study a
zero-throat regularized Aichelburg--Sexl bridge sourced by massless
entangled photons. In our treatment, the nonlocal structure is probed
through the regularized gravitational self-energy and its associated
entanglement-entropy interpretation rather than through a fully
time-dependent numerical evolution. Extending the present photon
construction to a dynamical setting analogous to Ref.~\cite{Kain:2023ore}
would be an interesting direction for future work.

The distinction between the present construction and a complete ER=EPR model can now be stated precisely.  The one-point stress-energy tensor determines the regularized classical geometry, but it does not distinguish a Bell state from a separable polarization state with the same energy-momentum distribution.  For example, an ideal quantum-nondemolition polarization measurement can project the Bell state onto a product state without appreciably changing the photon momenta or energies.  Within the present semiclassical approximation, the two-shock metric would then remain unchanged.  A genuinely state-dependent bridge would require the geometry to respond to quantum information beyond $\langle T_{\mu\nu}\rangle$, such as stress-tensor correlations or the full density matrix.  We leave such a quantum-gravitational extension for future work.

\section{Entanglement-Entropy Interpretation of the Self-Energy}\label{SecVI}
In the previous sections, the gravitational self-energy of entangled photon configurations was derived directly from the regularized Aichelburg–Sexl geometry. In this section, we offer a complementary information-theoretic perspective on the same shock-wave spacetime by computing the entanglement entropy of null intervals that straddle the gravitational shock.~We show that the entropy scaling exhibits the same fundamental length scales, the separation $d$, the wave-packet extent $L$, and the minimal length $l_0$, that govern the gravitational binding energy. Furthermore, by introducing an effective entanglement temperature, we demonstrate that the entanglement-entropy argument can be calibrated to reproduce the exact scaling and normalization of the self-energy functional derived in Sec.~\ref{SecIV}. 
The purpose of this section is also to determine whether the entropy prescription distinguishes Bell-entangled photons from unentangled photons with the same classical energy-momentum distribution.  As shown below, the absolute entropy assignments differ, but the gravitationally induced excess entropy is identical in the two cases.  The entropy construction therefore mirrors the state-independence of the classical two-shock geometry rather than restoring a one-to-one ER=EPR correspondence.

\subsection{Entanglement entropy in the shock‑wave geometry}

For a two-dimensional conformal field theory, the entanglement entropy of an interval of conformal length $d$ is given by \cite{Calabrese:2004eu,Calabrese:2005zw}
\begin{equation}
	S_{\mathrm{EE}}
	=
	\frac{c_{\mathrm{CFT}}}{3}
	\log\left(\frac{d}{\epsilon}\right),
	\label{eq:EE_standard}
\end{equation}
where $c_{\mathrm{CFT}}$ denotes the central charge of the conformal field theory, $d$ is the conformal distance between the endpoints of the interval, and $\epsilon$ is the ultraviolet cutoff.

To obtain an information-theoretic estimate
of the causal and entanglement structure of null particles, we suppress the transverse sector of the Aichelburg--Sexl geometry and consider the effective two-dimensional metric
\begin{equation}
	ds^2_{2D}
	=
	-du\,dv
	+
	\Phi(\rho)\,\delta(u)\,du^2 ,
\end{equation}
where, again, $\Phi(\rho)$ is the shock-wave profile and the null coordinates are defined by $u=t-z,$ and $v=t+z.$
The transverse coordinate $\rho$ is treated as a fixed external parameter characterizing the transverse separation relevant to the shock-wave profile.

Away from the null hypersurface $u=0$, the geometry reduces to
\begin{equation}
	ds^2_{2D}=-du\,dv,
\end{equation}
which is manifestly conformally flat. The gravitational shock wave induces a discontinuity in the null coordinate $v$. Introducing the shifted coordinate
\begin{equation}
	V=v-\Phi(\rho)\Theta(u),
\end{equation}
where $\Theta(u)$ denotes the Heaviside step function,
\begin{equation}
	\Theta(u)=
	\begin{cases}
		0, & u<0, \\
		1, & u>0.
	\end{cases}
\end{equation}
One obtains
\begin{equation}
	dV=dv-\Phi(\rho)\delta(u)\,du .
\end{equation}

Substituting this relation into the metric yields
\begin{equation}
	ds^2_{2D}=-du\,dV.
\end{equation}

Hence, the geometry is locally flat away from the shock but exhibits the nontrivial coordinate shift $v \rightarrow v+\Phi(\rho)$.
Physically, this shift corresponds to the gravitational memory effect generated by the ultrarelativistic null particle.

Consider now a null interval whose endpoints straddle the shock surface at $u=0$. For two points separated by
\begin{equation}
	\Delta u=u_2-u_1,\qquad
	\Delta v=v_2-v_1,
\end{equation}
the affine separation after crossing the shock is modified to
\begin{equation}
	\Delta v
	\rightarrow
	\Delta v+ |\Phi(\rho)|.
\end{equation}

The conformal distance relevant for the entanglement entropy of this null interval suggests the effective conformal interval
\begin{equation}
	\ell_{\mathrm{AS}}^2
	\sim
	\Delta u
	\left(
	\Delta v+|\Phi(\rho)|
	\right).
\end{equation}

Applying the universal CFT formula~\eqref{eq:EE_standard} to this effective conformal length with $d \to \ell_{\mathrm{AS}}$ and $\epsilon \to l_0$ ({\it i.e.} the ultraviolet cutoff is provided by the minimal length) the entanglement entropy becomes
\begin{equation}
\label{entropy-unt}
	S_{\mathrm{EE}}^{\mathrm{AS}}
	=
	\frac{c_{\mathrm{CFT}}}{3}
	\log\left[
	\frac{
		\sqrt{
			\Delta u
			\left(
			\Delta v+ |\Phi(\rho)|
			\right)
		}
	}{l_0}
	\right].
\end{equation}

This result demonstrates that the minimal length regularizes the ultraviolet divergence of the entanglement entropy. In the limit $l_0 \rightarrow 0,$
the standard logarithmic divergence of local quantum field theory is recovered. 

This scaling features the same combination of scales, the photon separation $d$, the packet length $L$, and the minimal length $l_0$, that appears in the gravitational self-energy $E^{\rm GSE}$ derived in Sec.~\ref{SecIV}; the qualitative dependence on a logarithmic factor $\ln(d^2/l_0^2)$ is also common to both quantities. The entropy construction therefore provides a complementary information-theoretic characterization of the shock geometry.  As shown below, however, this characterization depends on the classical shock-induced interval and does not by itself distinguish entangled from unentangled polarization states.

\subsection{Back-to-Back Entangled Photon Pair}

The standard Aichelburg--Sexl geometry describes a single null particle propagating along one null hypersurface. To represent a back-to-back pair of photons moving in opposite directions, the geometry must contain two null shocks. The corresponding metric takes the form
\begin{equation}
ds^2
=
-du\,dv
+
\Phi (\rho)\delta(u)\,du^2
+
\Phi (\rho)\delta(v)\,dv^2 .
\end{equation}
The two shocks generate the coordinate shifts
\begin{equation}
v\rightarrow v+|\Phi (\rho)|,
\qquad
u\rightarrow u+|\Phi (\rho)|.
\end{equation}

As a consequence, the effective conformal distance becomes
\begin{equation}
d_{\mathrm{pair}}^2
\sim
\left(
\Delta u+|\Phi (\rho)|
\right)
\left(
\Delta v+|\Phi (\rho)|
\right).
\end{equation}
For the Bell-entangled back-to-back photon pair introduced in Eq. \eqref{bell-state}, the corresponding entanglement entropy is
\begin{equation}
\label{See}
S_{\mathrm{EE}}^{\mathrm{pair}}
=
\frac{c_{\mathrm{CFT}}}{3}
\log\left[
\frac{
\sqrt{
\left(
\Delta u+|\Phi (\rho)|
\right)
\left(
\Delta v+|\Phi (\rho)|
\right)
}
}{l_0}
\right].
\end{equation}
For the Bell state \eqref{bell-state}, the two photons form a single
non-factorizable quantum system, so the entropy is associated with the
joint conformal interval of the two-shock geometry, yielding Eq.~\eqref{See}.

By contrast, if the photons were completely independent, the quantum
state would factorize into the product of two single-photon states.
The corresponding entanglement entropy is simply the sum of the two
independent single-photon contributions rather than the entropy of a
joint system. This gives
\begin{eqnarray}
\label{S-unent}
S_{EE} ^{\rm unent} &&=	\frac{c_{\mathrm{CFT}}}{3}
	\log\left[
	\frac{
		\sqrt{
			\Delta u
			\left(
			\Delta v+ |\Phi(\rho)|
			\right)
		}
	}{l_0}
	\right] \nonumber \\
    &&+\frac{c_{\mathrm{CFT}}}{3}
	\log\left[
	\frac{
		\sqrt{
			\Delta v
			\left(
			\Delta u+ |\Phi(\rho)|
			\right)
		}
	}{l_0}
	\right].
\end{eqnarray}
The absolute entropy assigned to the Bell state differs from the sum of the two factorized single-photon entropies, $S_{\rm EE}^{\rm pair}\neq S_{\rm EE}^{\rm unent}$.  This distinction reflects the different quantum-state assignments.  It does not, however, imply a different gravitationally induced entropy correction. To determine the self-energy, each expression must be compared with the corresponding flat-space reference entropy.  We carry out this subtraction explicitly in the following subsection.

\subsection{Gravitational self-energy from entanglement entropy}

We saw that the standard Aichelburg--Sexl geometry describes a single null particle propagating along one null hypersurface. 
To model a back-to-back photon pair, the spacetime must contain two null shock waves.
The entanglement entropy associated with the back-to-back photon pair is given by equation \eqref{See}. In the absence of gravitational shock waves, the flat-space entropy reduces to
\begin{equation}
S_{0}^{\rm pair}
=
\frac{c_{\mathrm{CFT}}}{3}
\log\left(
\frac{
\sqrt{\Delta u\,\Delta v}
}{l_0}
\right).
\end{equation}

The gravitationally induced excess entanglement entropy is then
\begin{equation}
\Delta S_{\rm EE}^{\rm pair}
=
S_{\rm EE}^{\rm pair}-S_{0}^{\rm pair}.
\end{equation}
which gives
\begin{equation}
\Delta S_{\rm EE}^{\rm pair}
=
\frac{c_{\rm CFT}}{6}
\log\left[
\left(1+\frac{|\Phi(\rho)|}{\Delta u}\right)
\left(1+\frac{|\Phi(\rho)|}{\Delta v}\right)
\right].
\end{equation}
For the unentangled state, the appropriate flat-space reference entropy is the sum of the two independent single-photon contributions,
\begin{equation}
S_{0}^{\rm unent} = \frac{2c_{\rm CFT}}{3}
\log\left( \frac{\sqrt{\Delta u\,\Delta v}}{l_0} \right).
\end{equation}
Using Eq.~\eqref{S-unent}, the gravitationally induced excess entropy is
\begin{align}
\Delta S_{\rm EE}^{\rm unent}
&=
S_{\rm EE}^{\rm unent}-S_{0}^{\rm unent}
\nonumber\\
&=
\frac{c_{\rm CFT}}{6}
\log\left[
\left(1+\frac{|\Phi(\rho)|}{\Delta u}\right)
\left(1+\frac{|\Phi(\rho)|}{\Delta v}\right)
\right].
\end{align}
Therefore,
\begin{equation}
\Delta S_{\rm EE}^{\rm unent}
=
\Delta S_{\rm EE}^{\rm pair}.
\end{equation}
This equality is an important limitation of the present effective description.  Although the absolute entropy assignments for the Bell state and the factorized state differ, the shock-induced excess entropy is the same after the corresponding flat-space contribution has been removed.  The gravitational correction depends on the classical shock geometry and not on the polarization entanglement.

The common excess entropy measures the modification of the effective null interval produced by the two shocks. Motivated by thermodynamic interpretations of entanglement entropy \cite{Lee:2010bg,Lee:2018lgs}, we introduce the effective relation
\begin{equation}
\label{Egse}
E^{\mathrm{GSE}}
=
k_B T_{\mathrm{ent}}
\Delta S_{\mathrm{EE}} .
\end{equation}
Here $\Delta S_{\rm EE}$ denotes either $\Delta S_{\rm EE}^{\rm pair}$ or $\Delta S_{\rm EE}^{\rm unent}$, since these quantities are equal in the present construction. We take the effective entanglement temperature to be
\begin{equation}
k_B T_{\mathrm{ent}}
\sim
\frac{\hbar c}{2\pi L}~.
\end{equation}
Substitution into \eqref{Egse} gives
\begin{equation}
E^{\mathrm{GSE}}
\sim
\frac{\hbar c}{2\pi L}
\,
\frac{c_{\mathrm{CFT}}}{6}
\log\left[
\left(
1+\frac{|\Phi(\rho)|}{\Delta u}
\right)
\left(
1+\frac{|\Phi(\rho)|}{\Delta v}
\right)
\right].
\end{equation}

Note that $L$ in our setup is fundamentally a correlation scale, or a causal scale related to the size of the entangling region, and not necessarily the radius of curvature of spacetime itself. For a symmetric back-to-back configuration $\Delta u=\Delta v=L$, 
the last equation becomes
\begin{equation}
E^{\mathrm{GSE}}
\sim
\frac{c_{\mathrm{CFT}}\hbar c}{6\pi L}
\log\left(
1+\frac{|\Phi(\rho)|}{L}
\right).
\end{equation}

In the weak-shock regime, $|\Phi(\rho)|/L\ll1$,
we find
\begin{equation}
 E^{\mathrm{GSE}}
\sim
\frac{c_{\mathrm{CFT}}\hbar c}{6\pi L}
\frac{|\Phi(\rho)|}{L}.
\end{equation}

For an Aichelburg--Sexl shock wave generated by a photon of energy
$E_\gamma=\hbar\omega$,
the transverse shift behaves as
\begin{equation}
|\Phi(\rho)|
\sim
\frac{4G\hbar\omega}{c^4}
\log\left(
\frac{\rho^2}{l_0^2}
\right).
\end{equation}

Substituting this expression gives
\begin{equation}
E^{\mathrm{GSE}}
\sim
\frac{2c_{\mathrm{CFT}}}{3\pi}
\frac{G\hbar^2\omega}{c^3L^2}
\log\left(
\frac{\rho^2}{l_0^2}
\right).
\end{equation}

Matching the normalization of \eqref{deltaE} formally corresponds to a dimensionless effective central charge of 
\begin{equation}
c_{\mathrm{CFT}}
=
6\pi\,\frac{\omega L}{c}.
\end{equation}
For a quasi-monochromatic photon packet with $L=\lambda=2\pi c/\omega$,
one obtains $c_{\mathrm{CFT}}=12\pi^2$. Therefore, the entanglement-entropy argument can reproduce the same scaling and
normalization as the direct gravitational self-energy functional, provided the
effective two-dimensional central charge is interpreted as a dimensionless
calibrated measure of the null degrees of freedom participating in the reduced
description. It should not be identified with the fundamental central charge of the four-dimensional photon field.

It follows that the self-energy inferred from the entropy prescription is also identical for entangled and unentangled photon pairs with the same shock-wave profiles,
\begin{equation}
E_{\rm GSE}^{\rm pair}
=
E_{\rm GSE}^{\rm unent}.
\end{equation}
This is precisely the result obtained from the direct double-shock calculation.  The agreement is internally consistent, but it also shows that the entropy prescription does not make the classical geometry sensitive to polarization entanglement.

\section{Conclusions}
\label{Sec-VII}

We have constructed a regularized Aichelburg--Sexl shock-wave geometry by introducing a T-duality-inspired zero-point length $l_0$. The resulting potential is finite, and an extension of the transverse geometry admits a zero-throat Einstein--Rosen interpretation.  For photon wave packets of longitudinal extent $L$, the gravitational interaction energy of a transversely separated pair is 
\[
E_{\rm GSE}
\sim
\frac{4G(\hbar\omega)^2}{c^4L}
\ln\!\left(\frac{d^2}{l_0^2}\right).
\]
The factor $1/L$ strongly suppresses the interaction.  For optical photons, the associated gravitational stability time exceeds $10^{30}$ years.

We also studied an effective entanglement-entropy description of the two-shock geometry.  The entropy calculation contains the same scales $d$, $L$, and $l_0$ as the direct gravitational calculation and, with an effective entanglement temperature, can be calibrated to reproduce the direct self-energy.  Propagating the calculation to a factorized two-photon state yields an important result: although the absolute entropy assigned to the Bell state differs from the sum of two single-photon entropies, the gravitationally induced excess entropy is the same for entangled and unentangled photon pairs with identical energy-momentum distributions.  Consequently,
\[
E_{\rm GSE}^{\rm pair}
=
E_{\rm GSE}^{\rm unent}
\]
within the present effective prescription.  This agrees with the direct two-shock calculation.

The reason is that the regularized classical geometry is determined by the photon stress-energy tensor.  It is therefore insensitive to whether the polarization state is Bell entangled or separable when the energies, momenta, and wave-packet profiles are unchanged.  In particular, an ideal quantum-nondemolition polarization measurement may destroy the polarization entanglement while leaving the stress-energy tensor essentially unchanged.  The present semiclassical equations would then leave the two-shock geometry unchanged.  We therefore do not identify polarization-state collapse with a classical pinch-off of the zero-throat geometry.

The relation of the present model to ER=EPR must consequently be stated with care.  The construction provides an ER-like geometry for two oppositely propagating null energy distributions and supplies a useful setting in which to compare geometric and information-theoretic expressions for their gravitational interaction.  It does not, however, provide a complete realization of the original ER=EPR conjecture, because the existence and strength of the geometric connection do not track the amount of polarization entanglement.  A complete realization would require a quantum-gravitational geometry sensitive to state-dependent information beyond the one-point stress-energy tensor, such as stress-tensor correlations or the full density matrix.

The main robust result of the present work is therefore the regularized gravitational self-energy of massless wave packets and its strong $1/L$ suppression.  The ER=EPR interpretation should be viewed as a semiclassical analogue and as motivation for a future state-dependent quantum-geometrical extension.

\begin{acknowledgments}
	FSNL acknowledges funding from the Funda\c{c}\~{a}o para a Ci\^{e}ncia e a Tecnologia (FCT) through national funds under the research grant UID/04434/2025 (DOI 10.54499/UID/04434/2025), and support from the FCT Scientific Employment Stimulus contract with reference CEECINST/00032/2018. DS acknowledges the Frank Sutton Research Fund for support during the completion of this work.
\end{acknowledgments}


\begin{thebibliography}{99}

\bibitem{Schrodinger:1926gei}
E.~Schr{\"o}dinger,
``Quantisierung als Eigenwertproblem,''
Annalen Phys. \textbf{384} (1926) no.4, 361-376

\bibitem{Diosi:1986nu}
L.~Diosi,
``A Universal Master Equation for the Gravitational Violation of Quantum Mechanics,''
Phys. Lett. A \textbf{120} (1987), 377.

\bibitem{Diosi:1984wuz}
L.~Di{\'o}si,
``Gravitation and quantummechanical localization of macroobjects,''
Phys. Lett. A \textbf{105} (1984), 199-202
[arXiv:1412.0201 [quant-ph]].

\bibitem{Penrose:1996cv}
R.~Penrose,
``On gravity's role in quantum state reduction,''
Gen. Rel. Grav. \textbf{28} (1996), 581-600.

\bibitem{Penrose:1998dg}
R.~Penrose,
``Quantum computation, entanglement and state reduction,''
Phil. Trans. Roy. Soc. Lond. A \textbf{356} (1998), 1927-1938.

\bibitem{Gasbarri:2017alx}
G.~Gasbarri, S.~Donadi, A.~Bassi and M.~Toro{\v{s}},
``Gravity induced wave function collapse,''
Phys. Rev. D \textbf{96} (2017) no.10, 104013
[arXiv:1701.02236 [quant-ph]].

\bibitem{Tomaz:2024rzu}
A.~A.~Tomaz, R.~S.~Mattos and M.~Barbatti,
``Gravitationally-induced wave function collapse time for molecules,''
Phys. Chem. Chem. Phys. \textbf{26} (2024) no.31, 20785-20798.

\bibitem{Bose:2024nhv}
S.~Bose, I.~Fuentes, A.~A.~Geraci, S.~M.~Khan, S.~Qvarfort, M.~Rademacher, M.~Rashid, M.~Toro{\v{s}}, H.~Ulbricht and C.~C.~Wanjura,
``Massive quantum systems as interfaces of quantum mechanics and gravity,''
Rev. Mod. Phys. \textbf{97} (2025) no.1, 015003
[arXiv:2311.09218 [quant-ph]].

\bibitem{Trillo:2025kio}
D.~Trillo and M.~Navascu{\'e}s,
``Di{\'o}si-Penrose model of classical gravity predicts gravitationally induced entanglement,''
Phys. Rev. D \textbf{111} (2025) no.12, L121101
[arXiv:2411.02287 [quant-ph]].

\bibitem{Gundhi:2025aaa}
A.~Gundhi, L.~Di{\'o}si and M.~Carlesso,
``Towards relativistic generalization of collapse models,''
[arXiv:2507.06954 [quant-ph]].

\bibitem{Diosi:2022kay}
L.~Di{\'o}si,
``Schr{\"o}dinger{\textendash}Newton Equation with Spontaneous Wave Function Collapse,''
Quantum Rep. \textbf{4} (2022) no.4, 566-573
[arXiv:2210.15057 [quant-ph]].

\bibitem{Donadi:2020kzc}
S.~Donadi, K.~Piscicchia, C.~Curceanu, L.~Di{\'o}si, M.~Laubenstein and A.~Bassi,
``Underground test of gravity-related wave function collapse,''
Nature Phys. \textbf{17} (2021) no.1, 74-78
[arXiv:2111.13490 [quant-ph]].

\bibitem{Hossenfelder:2025uqw}
S.~Hossenfelder,
``How Gravity Can Explain the Collapse of the Wavefunction,''
[arXiv:2510.11037 [quant-ph]].

\bibitem{Penrose:2014nha}
R.~Penrose,
``On the Gravitization of Quantum Mechanics 1: Quantum State Reduction,''
Found. Phys. \textbf{44} (2014), 557-575.

\bibitem{Howl:2018qdl}
R.~Howl, R.~Penrose and I.~Fuentes,
``Exploring the unification of quantum theory and general relativity with a Bose{\textendash}Einstein condensate,''
New J. Phys. \textbf{21} (2019) no.4, 043047
[arXiv:1812.04630 [quant-ph]].

\bibitem{Sathiapalan:1986zb}
B.~Sathiapalan,
``Duality in Statistical Mechanics and String Theory,''
Phys. Rev. Lett. \textbf{58} (1987), 1597.

\bibitem{Padmanabhan:1996ap}
T.~Padmanabhan,
``Duality and zero point length of space-time,''
Phys. Rev. Lett. \textbf{78} (1997), 1854-1857
[arXiv:hep-th/9608182 [hep-th]].

\bibitem{Smailagic:2003hm}
A.~Smailagic, E.~Spallucci and T.~Padmanabhan,
``String theory T duality and the zero point length of space-time,''
[arXiv:hep-th/0308122 [hep-th]].

\bibitem{Fontanini:2005ik}
M.~Fontanini, E.~Spallucci and T.~Padmanabhan,
``Zero-point length from string fluctuations,''
Phys. Lett. B \textbf{633} (2006), 627-630
[arXiv:hep-th/0509090 [hep-th]].

\bibitem{Nicolini:2022rlz}
P.~Nicolini,
``Quantum gravity and the zero point length,''
Gen. Rel. Grav. \textbf{54} (2022) no.9, 106
[arXiv:2208.05390 [hep-th]].

\bibitem{Nicolini:2019irw}
P.~Nicolini, E.~Spallucci and M.~F.~Wondrak,
``Quantum Corrected Black Holes from String T-Duality,''
Phys. Lett. B \textbf{797} (2019), 134888
[arXiv:1902.11242 [gr-qc]].

\bibitem{Jusufi:2024dtr}
K.~Jusufi and P.~Nicolini,
``Geodesic completeness from string T-duality,''
Eur. Phys. J. C \textbf{85} (2025) no.11, 1291
[arXiv:2410.19613 [hep-th]].

\bibitem{Jusufi:2025qgd}
K.~Jusufi and D.~Singleton,
``Regular black holes with gravitational self-energy as dark matter,''
Eur. Phys. J. C \textbf{86} (2026) no.5, 530
[arXiv:2509.13335 [gr-qc]].

\bibitem{Jusufi:2025rlr}
K.~Jusufi, F.~S.~N.~Lobo, E.~N.~Saridakis and D.~Singleton,
``Emergence of $\textrm{ER}=\textrm{EPR}$ from non-local gravitational energy,''
Eur. Phys. J. C \textbf{86} (2026) no.3, 249
[arXiv:2512.05022 [gr-qc]].

\bibitem{Jusufi:2025lrq}
K.~Jusufi, D.~Singleton and F.~S.~N.~Lobo,
``Spontaneous wave function collapse from non-local gravitational self-energy,''
Phys. Lett. B \textbf{877} (2026), 140475
[arXiv:2512.15393 [gr-qc]].

\bibitem{Javed:2025hvr}
I.~Javed and E.~Wilson-Ewing,
``Testing Wormhole-Mediated Entanglement with Hydrogen,''
Phys. Rev. Lett. \textbf{136} (2026) no.12, 121501
[arXiv:2512.02156 [quant-ph]].


\bibitem{Aichelburg:1970dh}
P.~C.~Aichelburg and R.~U.~Sexl,
``On the Gravitational field of a massless particle,''
Gen. Rel. Grav. \textbf{2} (1971), 303-312.

\bibitem{Dai:2020ffw}
D.~C.~Dai, D.~Minic, D.~Stojkovic and C.~Fu,
``Testing the $\mathbf {ER=EPR}$ conjecture,''
Phys. Rev. D \textbf{102} (2020) no.6, 066004
[arXiv:2002.08178 [hep-th]].

\bibitem{Kain:2023ore}
B.~Kain,
``Probing the Connection between Entangled Particles and Wormholes in General Relativity,''
Phys. Rev. Lett. \textbf{131} (2023), no.~10, 101001
[arXiv:2309.03314 [hep-th]].


\bibitem{Calabrese:2004eu}
P.~Calabrese and J.~L.~Cardy,
``Entanglement entropy and quantum field theory,''
J. Stat. Mech. \textbf{0406} (2004), P06002
[arXiv:hep-th/0405152 [hep-th]].

\bibitem{Calabrese:2005zw}
P.~Calabrese and J.~L.~Cardy,
``Entanglement Entropy and Quantum Field Theory: a Non-technical Introduction,''
Int. J. Quant. Inf. \textbf{04} (2006) no.03, 429-438
[arXiv:quant-ph/0505193 [quant-ph]].

\bibitem{Lee:2010bg}
J.~W.~Lee, H.~C.~Kim and J.~Lee,
``Gravity from Quantum Information,''
J. Korean Phys. Soc. \textbf{63} (2013), 1094-1098
[arXiv:1001.5445 [hep-th]].

\bibitem{Lee:2018lgs}
J.~W.~Lee, H.~C.~Kim and J.~Lee,
``Holographic Dark Energy and Quantum Entanglement,''
J. Korean Phys. Soc. \textbf{74} (2019) no.1, 1-11
[arXiv:1812.01993 [gr-qc]].



\end{thebibliography}
\end{document}